\documentclass[aps,prb,twocolumn,superscriptaddress]{revtex4}

\usepackage{graphicx}
\usepackage{dcolumn}

\begin{document}


\title{Dilute magnetism and vibrational entropy in Fe$_{2}$Al$_{5}$}


\author{Ji Chi}
\affiliation{Department of Physics, Texas A\&M University, College
Station, Texas 77843-4242, USA}
\author{Xiang Zheng}
\affiliation{Department of Physics, Texas A\&M University, College
Station, Texas 77843-4242, USA}
\affiliation{Materials Science and Engineering Program, Texas A\&M University, College
Station, Texas 77843-3003, USA}
\author{Sergio Y. Rodriguez}
\affiliation{Department of Physics, Texas A\&M University, College
Station, Texas 77843-4242, USA}
\author{Yang Li}
\affiliation{Department of Engineering Science and Materials, University of 
Puerto Rico at Mayaguez, Mayaguez, Puerto Rico 00681-9044, USA}
\author{Weiping Gou}
\author{V. Goruganti}
\author{K. D. D. Rathnayaka}
\author{Joseph H. Ross, Jr.}
\affiliation{Department of Physics, Texas A\&M University, College
Station, Texas 77843-4242, USA}
\affiliation{Materials Science and Engineering Program, Texas A\&M University, College
Station, Texas 77843-3003, USA}


\date{\today}

\begin{abstract}
Fe$_{2}$Al$_{5}$ contains a Fe-Al matrix through which are threaded
disordered one-dimensional chains of overlapping Al sites.
We report magnetic, NMR, and specific heat measurements
addressing its magnetic and vibrational properties.
The Curie-type susceptibility is found to be due to dilute moments, likely due to
wrong-site Fe atoms. $^{27}$Al NMR shift and spin-lattice relaxation
measurements confirm these to be indirectly coupled through an
RKKY-type interaction. Specific heat results indicate a large 
density of low-energy vibrational modes. These excitations 
generate a linear-$T$ contribution to the specific heat, which however freezes
out below about 10 K. These results are attributed to the presence of anharmonic
vibrational modes associated with the disordered structural chains.
\end{abstract}

\maketitle

\section {INTRODUCTION}
\begin{figure}
\includegraphics[width=\columnwidth]{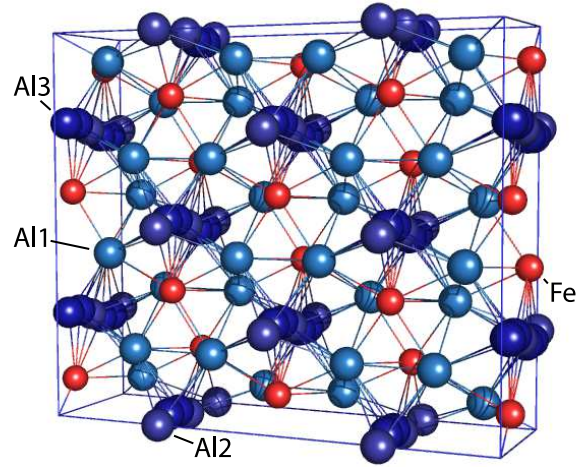}
\caption{\label{fig:fig1} [Color online] Fe$_{2}$Al$_{5}$ orthorhombic 
structure, showing four adjacent unit cells. Larger spheres denote Al, 
and smaller spheres Fe, as labeled. 
Undulating chains of overlapping Al(2) and Al(3) sites are shown here as if completely
occupied.}
\end{figure}

Al-rich Fe aluminides, along with other transition metal aluminides, form complex atomic structures,
for example the triclinic structure of FeAl$_{2}$,\cite{corby73}
and monoclinic structure of Fe$_{4}$Al$_{13}$, the latter a decagonal
quasicrystal approximant.\cite{grin94} In addition such
compounds show a variety of magnetic behavior, from non-magnetic ordered FeAl and dilute-Fe alloys,
\cite{guenzburger91,gonzales98,watson98} to concentrated-moment behavior in
FeAl$_{2}$.\cite{lue01,chi05}
Fe$_{2}$Al$_{5}$ is a quasicrystal approximant with pentagonal channels 
(Fig.~\ref{fig:fig1}), through which are threaded one-dimensional arrays
of partially-occupied aluminum sites spaced very closely and exhibiting\
anomalous vibrational properties.\cite{burkhardt94} In this paper we 
explore the magnetic and thermodynamic behavior of this material, showing
that it is an intrinsic dilute magnetic material with an anomalous contribution to
the entropy at low temperatures.

In some ways, these transition-metal aluminides may be viewed as dilute alloys
of $d$-rank ions in a metallic aluminum matrix. However, 
the complex structures also include regularly-spaced vacancies
and partially-occupied sites, stabilized in part by a combination
of strong $sp$-$d$ hybridization and the Hume-Rothery
mechanism.\cite{laissardiere05} In many cases this leads to
the appearance of a pseudogap in the density of electronic states at the Fermi level
[$g(\varepsilon_{F})$], and associated unusual electronic behavior. Furthermore, it is believed that the 
structures
of quasicrystals and approximants may lead to particularly
narrow structures in the DOS near 
$E_{F}$,\cite{fujiwara91,delahaye03,widmer09} 
which can dominate the electronic behavior. The possible
presence of phason-type coherent vibrational excitations\cite{coddens06,widom08}
is also of interest, and there is experimental evidence for an enhanced
density of vibrational modes in this class of materials.\cite{abe03,brand09}
Fe$_{2}$Al$_{5}$ also includes closely-spaced Al sites in the channels, and
hopping between these sites may have a large effect on the vibrational behavior.

The orthorhombic structure\cite{burkhardt94} (Cmcm, s.g. \#63) 
of Fe$_{2}$Al$_{5}$ is shown in Fig.~\ref{fig:fig1}. 
Its framework includes channels in the form of stacked
pentagonal antiprisms.  The single inequivalent Fe site
(4/cell) has two somewhat distant (3.06 \AA) Fe near neighbors, as well as eight Al(1) neighbors
and a series of neighboring channel-centered Al sites. 
Al(1) sites (8/cell) are fully occupied, while the overlapping
sites Al(2) (4/cell) and Al(3) (8/cell) channel sites make up disordered chains, 
with site occupation factors 0.36 and 0.23, respectively, found in the study
of Burkhardt {\it et al.}\cite{burkhardt94} 
These occupation factors  gives a nominal composition of Fe$_{4}$Al$_{10}$.
Reported composition ranges\cite{chen90,koster01} include variations as large as 
Fe$_{4}$Al$_{9}$ to Fe$_{4}$Al$_{11}$,
presumably through adjustment of Al(2) and Al(3) occupations.
The close spacing (0.67 and 0.75 \AA) of partially-occupied
channel sites and highly anisotropic thermal parameters deduced from 
crystallographic measurements also point to atomic hopping between
sites along the chains. A sizable Curie-type susceptibility was previously reported for
Fe$_{2}$Al$_{5}$,\cite{muller91} with $p_{eff}$ = 0.73. Here we report magnetization, 
specific heat and nuclear magnetic
resonance (NMR) studies addressing the magnetic and vibrational 
properties of this material.

\section{EXPERIMENTAL METHODS}

Samples were synthesized by arc melting appropriate
amounts of the elemental metals. The resulting polycrystalline 
ingots were annealed
in vacuum-sealed quartz tubes at 600$^{\circ}$ C for one week. 
Samples were characterized by powder x-ray
diffraction (Bruker D8 Advance) using Cu $K_{\alpha}$ radiation,
and also by wavelength dispersive x-ray spectrometry (WDS)
using a Cameca SX50 equipped with four 
wavelength-dispersive x-ray spectrometers. Structural
refinement of x-ray data was carried out using the GSAS software 
package.\cite{toby01,gsas} 

\begin{figure}
\includegraphics[width=\columnwidth]{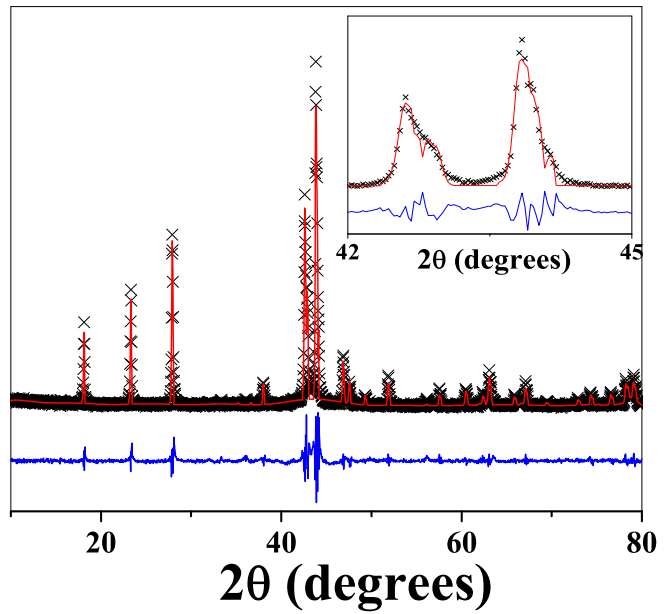}
\caption{\label{fig:fig2}Powder x-ray results for
Fe$_{2}$Al$_{5}$, with results of refinement and difference plot.}
\end{figure}

We prepared
samples with several starting compositions, including
Fe$_{4}$Al$_{10}$, Fe$_{4}$Al$_{11.2}$ and Fe$_{4}$Al$_{11.8}$.
WDS measurements showed 2 phases, FeAl$_{2}$ and
Fe$_{2}$Al$_{5}$, in the samples with Fe$_{4}$Al$_{10}$ and
Fe$_{4}$Al$_{11.2}$ starting compositions. However, for the starting
composition Fe$_{4}$Al$_{11.8}$ no second phase was found.
Also in the NMR measurements to be described
below, a separate peak due to the concentrated magnetic 
phase\cite{chi05} FeAl$_{2}$ was obtained for the
Fe$_{4}$Al$_{10}$ and Fe$_{4}$Al$_{11.2}$ samples, while a single peak was
found in Fe$_{4}$Al$_{11.8}$. WDS showed the final 
composition of the Fe$_{4}$Al$_{11.8}$
sample to be Fe$_{4}$Al$_{9.8}$ after loss of Al. 
X-ray diffraction results for this sample are shown in Fig.~\ref{fig:fig2}. Atomic
parameters obtained from the refinement are in reasonable agreement with those reported
earlier.\cite{burkhardt94} 
Unless otherwise noted, results reported here are for this sample.
Where necessary, samples are designated according to the aluminum starting 
composition, for example  sample Al$_{11.8}$.

Specific heat measurements were performed using a Quantum 
Design Physical Property Measurement System from 
room temperature to 1.8 K. DC-susceptibility measurements 
were carried out using  a Quantum
Design MPMS SQUID magnetometer. $^{27}$Al NMR experiments were performed 
from 4.2 K to 450 K at two fixed
fields using a wide-line pulse spectrometer. 
For these measurements the powdered sample was mixed with granular quartz, and 
aqueous AlCl$_{3}$ was used as shift reference except as noted. MAS-NMR
measurements were also performed at room temperature using an Advance 400
spectrometer.

\section{RESULTS AND ANALYSIS}

\subsection{Magnetic measurements}

\begin{figure}
\includegraphics[width=\columnwidth]{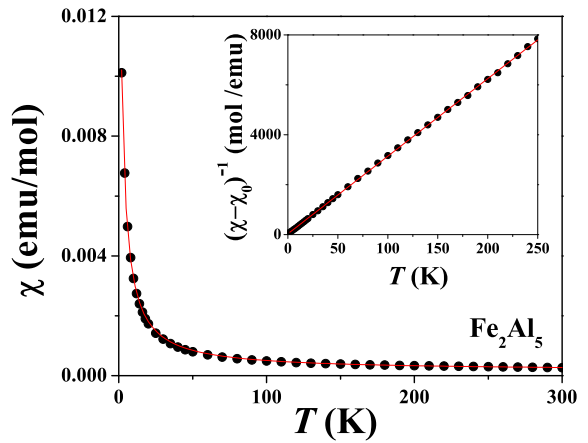}
\caption{\label{fig:fig3} Fe$_{2}$Al$_{5}$ DC susceptibility ($\chi(T)$) per mol
Fe in an applied field of 1000 Oe. Inset: $(\chi-\chi_{o})^{-1}$
vs. temperature. Solid curves: Curie-Weiss fit. }
\end{figure}

The dc-susceptibility, $\chi(T)$, is shown in Fig.~\ref{fig:fig3}, for fixed
field of 1000 Oe. 
We fit the data to a Curie-Weiss function according to the standard 
relationship,
$\chi(T)=C_{cw}/(T-\theta)+\chi_{o}$, with 
$\chi_{o}$ = + 1.7 $\times 10^{-4}$ emu/mol a small extrinsic 
contribution. The Curie constant is
\begin{equation}
C_{cw}=N_{A}c\frac{p_{eff}^{2}\mu_{B}^{2}}{3k_{B}} \label{eq:7},
\end{equation}
where $N_{A}$ is Avogadro's number, $c$ the concentration of
magnetic ions per iron atom, and $p_{eff}$ the effective moment. From
least-squares fits we obtained $\theta$ = $-$1.1 K, and $(p_{eff}\sqrt{c})$ =
0.51, somewhat smaller than the previous report,\cite{muller91} the difference likely
due to sample-dependent variations in atomic occupation parameters.

\begin{figure}
\includegraphics[width=\columnwidth]{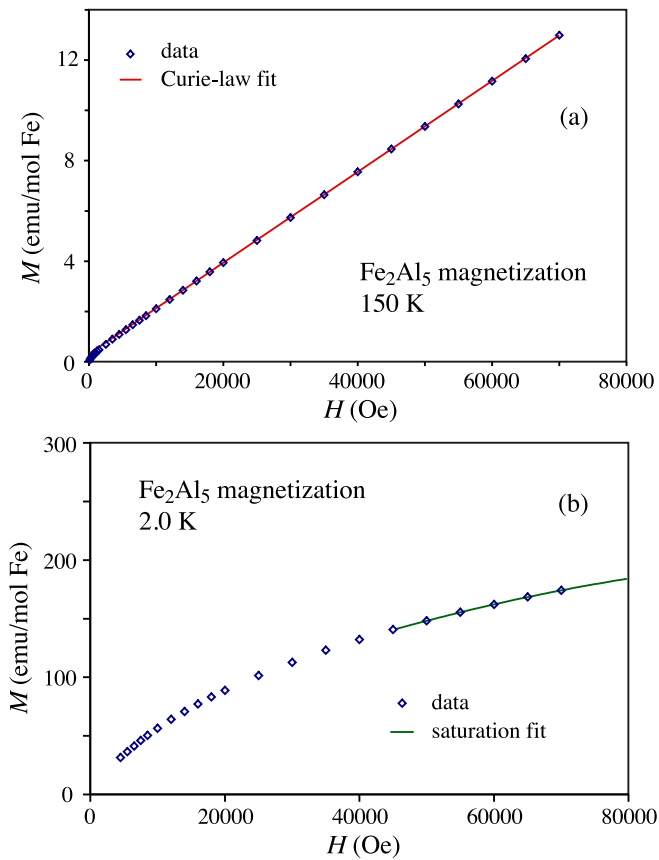}
\caption{\label{fig:figMH} Fe$_{2}$Al$_{5}$ magnetization per mol
Fe. (a) Measurements at 150 K with Curie-Weiss fit described in the text.
(b) 2.0 K, with saturation fit described in text.}
\end{figure}

Magnetization measurements vs. field at 150 K show linear behavior
[Fig.~\ref{fig:figMH}(a)]. The small vertical offset corresponds to
the extrinsic contribution $\chi_{o}$ described previously. 
The straight-line fit shown in Fig.~\ref{fig:figMH}(a)
yields $(p_{eff}\sqrt{c})$ = 0.47. 
At 2.0 K [Fig.~\ref{fig:figMH}(b)] the magnetization is 
nonlinear, but with incomplete saturation at the highest
field of 70 kOe. This curve could not be fitted to a Brillouin function 
representing the magnetizing behavior of independent local
moments, presumably due to spin interactions corresponding to the
small fitted $\theta$ obtained from the susceptibility. However,
the saturation could be fitted to a function\cite{Chikazumi97}
$M = M_{sat} - A/H - B/H^2$ [solid curve in Fig.~\ref{fig:figMH}(b)],
yielding $M_{sat}$ = 281 emu/mol Fe. Since for local moments $p_{eff} = g\sqrt{J(J + 1)}$ while the
saturation moment corresponds $gJ\mu_{B}$ per ion, $J$ and $c$ can 
be determined separately from
the $MH$ curves, with $g$ known. Assuming that $g$ = 2
(a reasonable assumption for transition ions), this procedure yields
$J$ = 1.2 (hence $p_{eff}$ = 3.2), and $c$ = 0.021 per Fe atom.

\subsection{NMR measurements}

\begin{figure}
\includegraphics[width=\columnwidth]{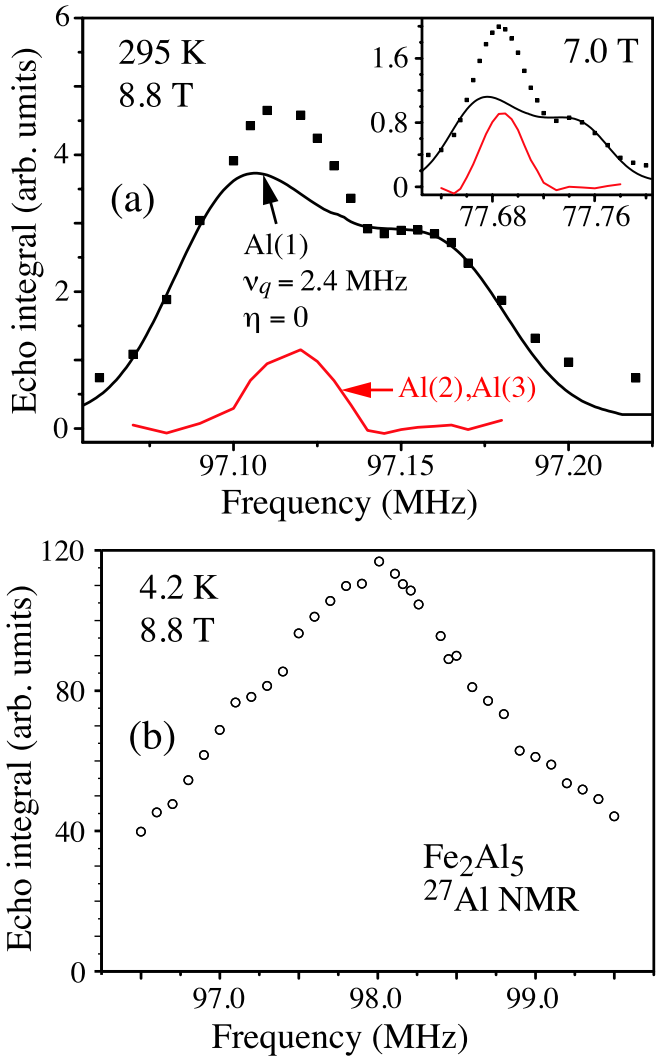}
\caption{\label{fig:figNMR} $^{27}$Al NMR lineshapes for Fe$_{2}$Al$_{5}$. 
(a) 295 K measurement, with calculated quadrupole powder pattern for Al(1) sites
(upper curve) and difference curve attributed to Al(2) and Al(3) sites (lower curve). 
Inset:  lower-field measurement, with curves calculated using same
parameters. (b) 4.2 K measurement.}
\end{figure}

\begin{figure}
\includegraphics[width=\columnwidth]{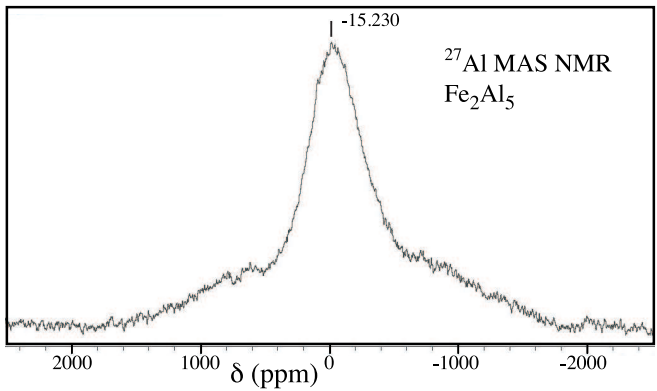}
\caption{\label{fig:figMAS} $^{27}$Al MAS-NMR spectrum for Fe$_{2}$Al$_{5}$, in 9.4 T.
Spinning rate = 12 kHz. Shift reference: Al$_{2}$O$_{3}$.}
\end{figure}

The $^{27}$Al NMR spectrum at room temperature has a two-peaked structure 
[Fig.~\ref{fig:figNMR}(a)], while at lower
temperatures the linewidth becomes progressively larger due to the
presence of paramagnetic moments [Fig.~\ref{fig:figNMR}(b)]. The pulse-length
behavior\cite{kanert71} confirms the central portions of these lines to be central transitions
for $^{27}$Al ($I$ = 5/2) connecting $-$1/2 $\longleftrightarrow$ +1/2 states.
A room-temperature MAS-NMR spectrum is shown in Fig.~\ref{fig:figMAS}. The observation
of a single peak in this graph indicates that the structure in the static line is dominated by orientation dependence
of a single site, rather than a splitting due to inequivalent sites, even though the random
paramagnetic contribution is too large to be effectively narrowed by spinning in this case. 
Since 8 of the nominal 10 Al atoms per cell belong to 
Al(1), we can deduce that the main features of the line correspond 
this site, with resonances due to the other sites either overlapping this line or
excessively broadened due to channel disorder.

The field-dependence enables magnetic and quadrupole 
contributions to be separated,\cite{stauss64,carter77} since 
the magnetic shift is proportional to
$B$ and the second-order quadrupole shift ($\Delta f_{Q}^{2}$) is proportional to $1/B$. Fitting the 295 K
spectral center of mass obtained at fixed fields of 7.00 and 8.75 T, we extracted a
mean magnetic shift, $K$ = +0.020(1)\%, and a corresponding mean quadrupole contribution
$\Delta f_{Q}^{2} B$ = $-$0.085 MHz T. The latter corresponds to the shift for a second-order
powder pattern having quadrupole parameters\cite{stauss64} $\nu_{Q}$ = 1.9 MHz and $\eta$ = 0
(or $\nu_{Q}$ as small as 1.6 MHz with $\eta$ increased to 1). 
However, in numerical calculations starting with these values we found that the main
features of the spectra could be fitted consistently vs. field only by using a larger $\nu_{Q}$,
with $\eta \approx$ 0 and $K$ = +0.025\%. Fig.~\ref{fig:figNMR}(a) shows the superposition of such a
single-site quadrupole spectrum with $\nu_{Q}$ = 2.39 MHz for two different fields. The results are relatively insensitive to the
parameter $\eta$, however we found that computed curves for $\eta \gtrsim$ 0.3 agreed poorly with the data.
With this plot superposed, we find in addition a second peak,
as shown from the difference curves plotted in Fig.~\ref{fig:figNMR}(a). This additional peak makes up
12--22\% of the total spectral weight. The difference in magnitude may be an artifact
of the fitting. According to the measured site occupations, 20\% the Al reside on the two channel sites,
so it is reasonable to conclude that the additional peak corresponds to these sites, while the fitted $\nu_{Q}$ = 2.39 MHz
resonance is attributed to Al(1).

\begin{figure}
\includegraphics[width=\columnwidth]{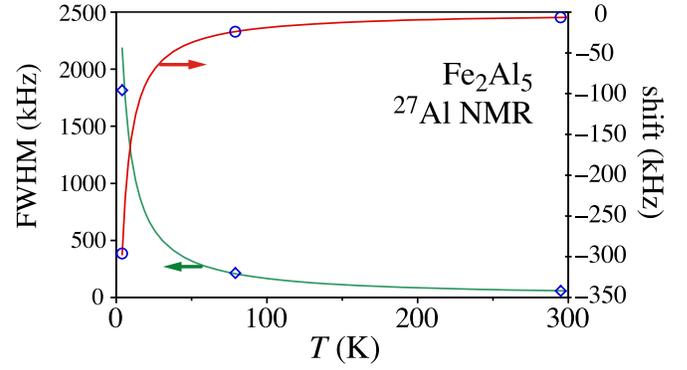}
\caption{\label{fig:figShifts} $^{27}$Al NMR line shifts and widths, plotted
with fitted Curie-Weiss curves described in text.}
\end{figure}

The low-temperature behavior of the NMR lines was measured in 8.8 T at 79 K and 4.2 K; the extracted 
shifts and linewidths are plotted in Fig.~\ref{fig:figShifts}. Widths are full widths at 
half maximum (FWHM) obtained through line fitting, with the quadrupole contribution removed, 
and shifts are those of the spectral
centers of mass. These data could be fit to Curie-Weiss functions, as plotted in the
figure. The fitted Weiss temperatures are $-$3.8 and $-$2.6 K, respectively, comparable
to the small $\theta$ obtained from the susceptibility. Therefore the local fields at these temperatures
are dominated by these paramagnetic moments. The nonzero average shift
indicates RKKY-type rather than simple dipolar nuclear coupling to the dilute paramagnetic
moments, since the dipolar coupling has zero spatial average. 

\begin{figure}
\includegraphics[width=\columnwidth]{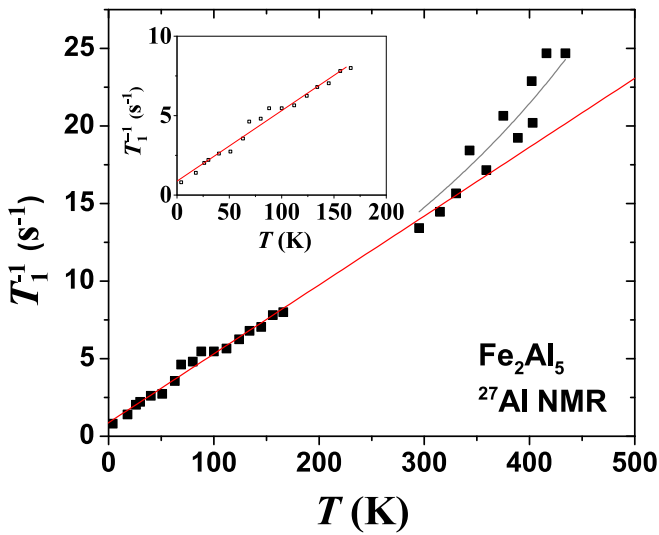}
\caption{\label{fig:fig4}Temperature dependence of relaxation
rates for $^{27}$Al, measured at 8.75 T. Solid line: fit to Korringa +
paramagnetic relaxation. Upper curve is guide to the eye. Inset: Low-temperature portion
of data.}
\end{figure}

$^{27}$Al NMR spin-lattice relaxation times ($T_{1}$'s) were
measured using the inversion recovery method, recording the
signal strength by integrating the spin echo FFT. 
For the central transitions, the $T_{1}$'s were extracted by fitting to
multiexponential curves \cite{narath67} for magnetic relaxation of the
$I$ = 5/2 $^{27}$Al central line. In Fig.~\ref{fig:fig4} we show the
temperature dependence between 4 and 450 K. These data were
measured in 8.8 T; measurements at selected temperatures also
indicated no field-dependence to the $T_{1}$.
The straight line in Fig.~\ref{fig:fig4} represents a fit to the data below
200 K of the form
\begin{equation}
T_{1}^{-1}(T)=aT+T_{1P}^{-1} \label{eq:1},
\end{equation}
with $a$ = 4.44 $\times$ 10$^{-2}$ K$^{-1}$s$^{-1}$, 
and $T_{1P}^{-1}$ = 0.87 s$^{-1}$. The first term in
Eq.~\ref{eq:1} follows the expected Korringa behavior
for metals. Subtracting the small negative
Curie-type shift (obtained from the fit shown in Fig.~\ref{fig:figShifts}, and 
amounting to $-$0.007\% at room temperature) from $K$ for the main
Al(1) site as described above, we obtain a baseline room-temperature metallic
shift, $K_{m}$ = 0.032\%. Multiplying by the fitted slope from
Eq.~\ref{eq:1} yields the Korringa product\cite{carter77}
$K^{2}(T_{1}T) \equiv K_{m}^{2}a^{-1} = 2.3 \times 10^{-6}$ sK. Compared to the
standard $^{27}$Al Korringa value, $3.9 \times 10^{-6}$ sK,
the smaller value obtained here likely indicates 
the presence of a negative cross-polarization term, and $d$ contributions
to $g(\varepsilon_{F})$ that will tend to
reduce $K_{m}$, and thereby the Korringa product.

The additive term $T_{1P}^{-1}$ in Eq.~\ref{eq:1} can result
from the presence of paramagnetic moments interacting 
through a constant RKKY-type coupling field. We can estimate such
a contribution by calculating the nuclear transfer hyperfine
coupling as follows: the fitted
paramagnetic contribution to the shift in Fig.~\ref{fig:figShifts}
at 79 K is $-$24 kHz, which corresponds to a 22 G effective
field for $^{27}$Al. Also at this temperature the paramagnetic
response will be linear; the parameters used for Fig.~\ref{fig:figMH}(a)
yield $M$ = 0.27 $\mu_{B}$ per paramagnetic ion at this
temperature and field, so the transfer field is 81 G/$z$, where 
$z$ is the effective coordination number for paramagnetic 
ions interacting with each nuclear spin. Dividing by the
nuclear gyromagnetic ratio we obtain the transfer coupling
parameter, $A/h$ = 89 kHz/$z$. The contribution of the paramagnetic
moments with spin $J$ to the $T_{1}$ can be expressed\cite{narath67}
\begin{equation}
T_{1P}^{-1} = (2\pi )^{-3/2}(A/h )^{2}(3\omega _{E})^{-1}J(J+1)z
\label{eq:three}.
\end{equation}
In this case $\omega _{E}$ is the average RKKY exchange frequency, which
may be expressed,
$[8J_{ex}^{2}zJ(J+1)/3\hbar ^{2}]^{1/2}$, where $J_{ex}$ is 
the interaction strength. In the paramagnetic regime \cite{ashcroft76},
$J_{ex}$ is related to the Curie temperature through $J_{ex}z =\frac{3k_{B}\theta}{2J(J + 1)}$. Combining, and using the
fitted $\theta$ = $-$1.1 K and $J$ = 1.2, we obtain $T_{1P}^{-1}$ = 2.1 s$^{-1}/\sqrt{z}$. Assuming the dilute
moments to be randomly positioned, the value for $z$ is unclear, however it should be a small number on order
of unity. Thus this estimate is in quite good agreement with
the fitted $T_{1P}^{-1}$ = 0.87 s$^{-1}$, providing additional
evidence supporting the picture of a nonmagnetic matrix
containing weakly coupled dilute moments. 

Above 300 K, there is an increase in $T_{1}^{-1}$ beyond the linear Korringa
fit. Similar behavior has been observed in a number of
quasicrystaline alloys\cite{dolinsek01,jeglic05}, and
attributed in some cases to a pseudogap near the Fermi
energy, whereby an increase in temperature leads to an increase
in the number of carriers available for Korringa relaxation.
However we see no evidence for a corresponding increase
in the Knight shift. While we have not mapped out the detailed
field-dependence of the lineshapes above 295 K, we measured lineshapes 
at 7.0 T at several temperatures up to 403 K. We found
no change in the center-of-mass central line position
between 295 and 403 K within experimental error. 
Thus a more likely explanation for the increase in $T_{1}^{-1}$
is vibrational. Indeed, there are a number of reports of enhanced
density of localized vibrational modes at high temperatures in quasicrystals
as well as in quasicrystal approximants,\cite{dolinsek02,widom08,brand09} and
in glassy systems,\cite{szeftel75,mammadov10} due to thermal excitation of localized vibrational
modes. For comparison, the $^{27}$Al $T_{1}^{-1}$ contribution (quadratic at high temperatures)
due to ordinary phonons in AlSb is roughly 10 times smaller.\cite{mieher62}

\subsection{Specific heat}

\begin{figure}
\includegraphics[width=\columnwidth]{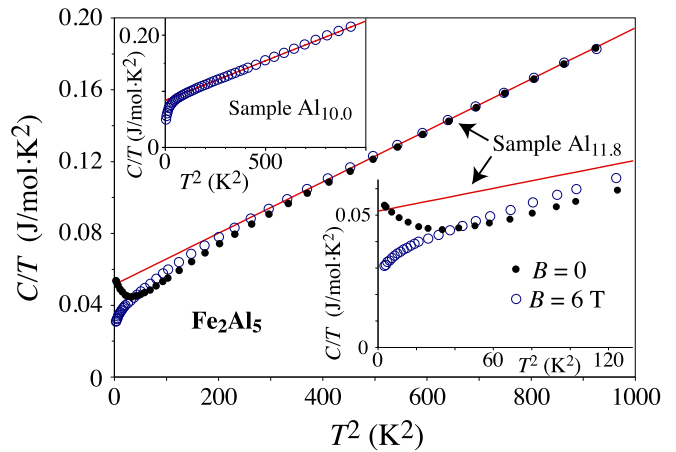}
\caption{\label{fig:fig5} $C/T$ vs. $T^{2}$ for Fe$_{2}$Al$_{5}$
between 0 and 30 K, for samples and fields as indicated. 
Calibration is per mole formula unit, Fe$_{2}$Al$_{5}$.
Solid lines correspond to the fitted parameters described in the text.}
\end{figure}

The specific heat ($C$) was measured in the temperature range 1.8--300 K. 
$C/T$ vs. $T^{2}$ plots below 30 K are shown in
Fig.~\ref{fig:fig5}, for the Al$_{11.8}$ and the Al$_{10.0}$ sample, the
latter including a small FeAl$_{2}$ minority phase. In both cases the
straight-line behavior is equivalent to that of typical metals, 
however with a downturn at low temperatures. 
A fit to $C(T) = \gamma T + \beta T^{3}$ for Al$_{11.8}$ yielded 
$\gamma$ = 48 mJ/mol K$^{2}$
and $\beta$ = 0.147 mJ/mol K$^{4}$ with zero applied field (solid line in
Fig.~\ref{fig:fig5}). Data with 6 T applied field are also shown;
a similar fit yields $\gamma$ = 52 mJ/mol K$^{2}$
and $\beta$ = 0.143 mJ/mol K$^{4}$. For sample Al$_{10.0}$
the results are $\gamma$ = 83 mJ/mol K$^{2}$
and $\beta$ = 0.141 mJ/mol K$^{4}$ (shown in inset plot).
The essentially identical $\beta$ values correspond 
to a Debye temperature $\Theta_{D}$ = 460 K, however in both cases the
fitted $\gamma$ is quite large for a nonmagnetic metal. The low-temperature
downturn is anomalous, and appears in both samples. 
Henceforth, we will focus on sample Al$_{11.8}$, on which the 
previously-described magnetic and NMR measurements were also performed.

\begin{figure}
\includegraphics[width=\columnwidth]{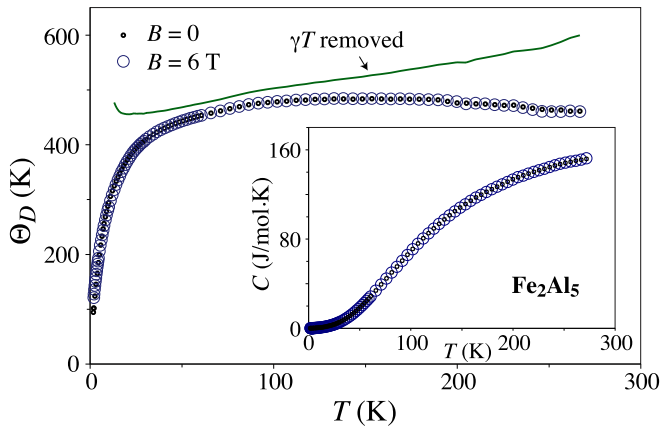}
\caption{\label{fig:fig_debye} Debye temperature extracted from  Fe$_{2}$Al$_{5}$
specific heat, for $B$ = 0 and 6 T. Solid curve: calculated
after removal of $\gamma T$ as described in text (6 T data). 
Inset:  specific heat measured in $B$ = 0 and 6 T, per mole formula unit.}
\end{figure}

The inset in Fig.~\ref{fig:fig_debye} shows the specific heat for the entire temperature range. On this scale the
6 T and 0 T data appear superposed. One can solve for the Debye temperature
by solving the standard integral equation, $C = 9NR(T/\Theta_{D})^{3}\int_0^{\Theta_{D}/T}x^{4}e^{x}dx/(e^{x}-1)^2$,
with $\Theta_{D}$ as a parameter.\cite{cezairliyan88} Assuming $C$
to be entirely vibrational, we obtain the lower curves  displayed in the main plot of Fig.~\ref{fig:fig_debye}. 
The large drop in $\Theta_{D}$ below 50 K signals the presence of additional
low-energy excitations in this temperature range, which can be associated with
the large $T$-linear term fitted as $\gamma T$. Subtracting
this term from the measured specific heat before solving for $\Theta_{D}$
yields the upper curve in Fig.~\ref{fig:fig_debye}, which brings $\Theta_{D}$
back into agreement with the value, 460 K, corresponding to the measured
$\beta T^{3}$. However, the continued rise of this curve as room temperature is approached
is unphysical and indicates that the large $\gamma T$ contribution does not extend to
high temperatures as would the electronic contribution in an ordinary metal. 
Instead, these additional excitations are apparently confined to a plateau in $C/T$.

\begin{figure}
\includegraphics[width=\columnwidth]{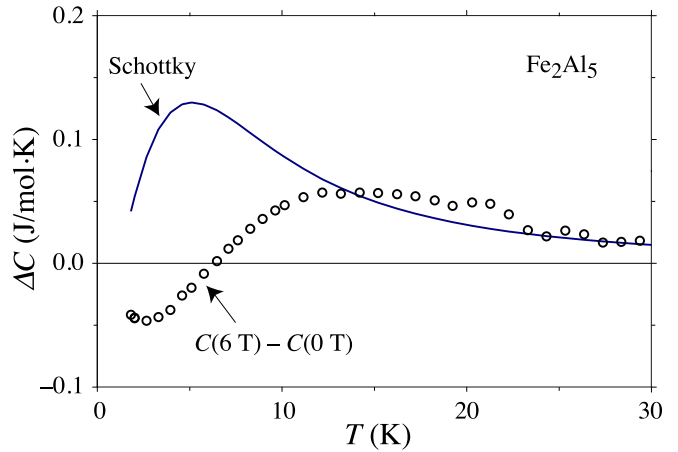}
\caption{\label{fig:fig_Cdiff} Specific
heat difference for zero and 6 T data, with calculated $B$ = 6 T multilevel Schottky anomaly
for fitted density of paramagnetic defects.
Normalization is per mole formula unit.}
\end{figure}

The small low temperature difference between specific heat curves for 6 T and 0 T 
is consistent with a Schottky anomaly for paramagnetic spins. The difference between the two measurements
is plotted in Fig.~\ref{fig:fig_Cdiff}, allowing
other contributions to be removed in a model-independent way. Also plotted is 
a multilevel Schottky curve\cite{cezairliyan88} for $B$ = 6 T corresponding to paramagnetic defects
with $g$ = 2 and $c$ = 0.021 as obtained from the magnetic measurements, and using
$J$ = 3/2 to approximate the fitted $J$ = 1.19. Above 10 K where the $B$ = 0 contribution has
apparently died out, the magnitudes are in agreement. The negative peak near 2 K
corresponds to the zero-field upturn in $C/T$ at low $T$, and is consistent with spin freezing  
due to a distribution of local fields. The contribution of the measured dilute moments
to the specific heat is described entirely by these low-temperature effects, and thus the
low-energy excitations corresponding to the plateau in $C/T$ 
must be attributed to nonmagnetic features. 

\begin{figure}
\includegraphics[width=\columnwidth]{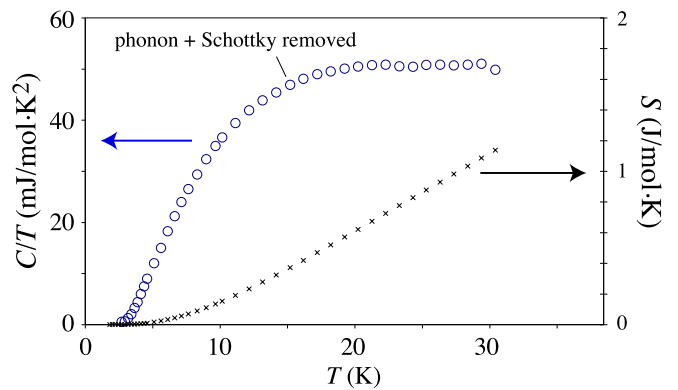}
\caption{\label{fig:fig_entropy} Reduced $C/T$ with $\beta$ $T^{2}$
and magnetic Schottky terms removed (left scale). Also plotted: entropy obtained from reduced $C/T$ (right scale).
Units are per mole formula unit.}
\end{figure}

With both the fitted $\beta T^{3}$ and Schottky terms removed, the remaining low-temperature
$C/T$ contribution is plotted in Fig.~\ref{fig:fig_entropy}. In typical metals this would yield a constant
equal to the electronic $\gamma$, however in this case the curve trends toward zero as
$T$ approaches zero. The Schottky term is likely overestimated, as $J$ = 3/2
was used as an approximation for the fitted $J$ = 1.2. Thus the remaining metallic $\gamma$
is likely nonzero, but it is clearly significantly smaller than the apparent $\gamma$ 
of the plateau above 20 K. The integrated entropy [$S = \int (C/T)dT$] is also shown,
obtained from the reduced $C/T$. This curve is still rising at 30 K (above which temperature
additional terms in the ordinary phonon expansion must be included), however the change in $S$
over these temperatures provides an estimate of the magnitude of the excess vibrational contribution.

\section{DISCUSSION}

From the NMR and magnetic measurements it is clear that this is an ordinary metal containing relatively
dilute paramagnetic defects. The size of these moments is comparable in size to that found for
thermally-generated Fe antisite defects\cite{Lue99} in Fe$_{2}$VAl.
In the present case a likely scenario may be that Fe atoms randomly substituting
on chain sites \cite{burkhardt94} 
may be magnetic due to the more closely connected Fe neighbors of these
sites. It is also possible that built-in chain-site disorder  
induces moments on a small fraction of the otherwise nonmagnetic Fe sites; generally
for dilute Al-Fe alloys the presence of a local moment is a sensitive
function of the coordinating atom positions.\cite{gonzales98,mantina09}
Furthermore, as has been proposed for Mn-aluminide 
quasicrystals\cite{laissardiere05} complex overlapping RKKY interactions in these aluminides can 
reinforce dilute magnetic moments on certain sites within the otherwise nonmagnetic material.

$K_{m}$, the baseline metallic shift as extracted here for the metallic matrix, is about 20\% of the value for Al metal. This is likely associated 
with a broad minimum in the Al partial $g(\varepsilon)$ near $\varepsilon_{F}$ as is typical of hybridization 
gap systems.\cite{watson98} However we found no evidence for a temperature dependence to $K_{m}$, as
has been found in some systems\cite{lue98,tang97} for which narrow pseudogap features dominate near $\varepsilon_{F}$. 
Thus from this standpoint Fe$_{2}$Al$_{5}$ behaves as an ordinary metal. Additional excitations identified
in this material take the form of localized modes excited at high temperatures, evidenced by an enhancement in
$T_{1}$, and low-energy features that dominate the specific heat, shown to be nonmagnetic in origin.

Low-energy Einstein oscillator-type vibrational modes have been observed
previously in metallic systems,\cite{caplin73,zhou06} however the low-temperature specific heat 
observed here is not consistent with a single oscillator with a well-defined frequency. 
On the other hand the alternative explanation of a narrow electronic pseudogap,
which could produce the low-temperature reduction in specific heat as observed here, seems unlikely since an extremely 
narrow gap would be required. In addition the observed field-independence would rule out magnetic Kondo 
behavior as a source of such a feature. 
Thus we 
conclude that the enhanced low-energy excitations are due to an extended spectrum of localized anharmonic vibrations. 
We attribute these to atomic motion along the unusual partially-occupied chains found in this material. 

An enhanced specific heat nearly linear in the temperature is observed in insulating 
glasses\cite{phillips87} as well as orientational glasses,\cite{talon02} and these are modeled as localized atomic hopping sites 
producing a broad spectrum of two-level systems. For this model to work in the present case, a low-energy 
cutoff must be present in the excitation spectrum, with an energy close to 1 meV corresponding to the observed
loss of excitation strength near 10 K. This cutoff could plausibly be attributed to 
the energy required for hopping between neighboring partially-occupied chain sites. 
The large linear coefficient of specific heat in both measured samples
implies a high density of these vibrational modes. For example the integrated
entropy shown in Fig.~\ref{fig:fig4} reaches 0.14 $R$ at 30 K; if attributed to two-level localized anharmonic states
this implies approximately 0.14/ln2 = 0.2 such modes per formula unit becoming activated below 30 K,
or a significant fraction of the chain sites, assuming that each localized vibrational state corresponds to a 
single chain atom. We assume that the breadth of the vibrational spectrum is be due to the disordered arrangement
on these sites. 

\section{CONCLUSIONS}

Magnetic measurements of Fe$_{2}$Al$_{5}$ show it to be a nonmagnetic metal with
a large density of paramagnetic moments, likely connected to wrong-site Fe atoms on the disordered
chains. $^{27}$Al NMR results confirm that these are indirectly coupled through an RKKY-type
interaction via the nonmagnetic matrix. We identified Korringa behavior in NMR shifts and relaxation
rates; an additional contribution to the latter was identified as due to localized vibrational excitations.
Specific heat measurements and analysis demonstrate in addition a large density of low-energy modes
contributing to a significant linear-$T$ enhancement in the specific heat. An anomalous freeze-out of these
modes is observed at low temperatures.
We showed these to be nonmagnetic, and likely attributable to atomic motion on the partially-occupied 
chains threading the structure of this material.

\begin{acknowledgments}

This work was supported by the Robert A. Welch Foundation, Grant
No. A-1526, by the National Science Foundation (DMR-0315476), and
by Texas A\&M University through the Telecommunications and
Informatics Task Force.

\end{acknowledgments}

\bibliography{Fe2Al5}

\end{document}